\documentclass{pasj00}

\draft

\begin{document}
\SetRunningHead{S. Kato}{}
\Received{2013/00/00}
\Accepted{2013/00/00}

\title{Resonant Excitation of Disk Oscillations in Deformed Disks VI: Stability Criterion Revisited} 

\author{Shoji \textsc{Kato}}
\affil{2-2-2 Shikanodai-Nishi, Ikoma-shi, Nara, 630-0114}
\email{kato@gmail.com, kato@kusastro.kyoto-u.ac.jp}

%


\maketitle

\begin{abstract} 

We re-examine excitation of a set of disk oscillations in a deformed disk by a resonant process.
We assume that the disk is deformed from an axisymmetric steady state by an oscillatory deformation with frequency $\omega_{\rm D}$
and azimuthal wavenumber $m_{\rm D}$.
Then, we consider two normal mode oscillations with a set of frequencies and azimuthal wavenumber being ($\omega_1$, $m_1$) and 
($\omega_2$, $m_2$) and satisfying the resonant conditions ($\omega_1+\omega_2+\omega_{\rm D}=0$ and
$m_1+m_2+m_{\rm D}=0$).
These oscillations  are resonantly excited if $(E_1/\omega_1)(E_2/\omega_2)>0$, 
where $E_1$ and $E_2$ are wave energies of the above two oscillations,
when the deformation is maintained by external forces or has a large amplitude compared with the oscillations.
This instability condition is rather general as long as unperturbed density and pressure vanish on the surface of the system.
Possibility of application to superhump and negative superhump in superoutburst state of dwarf novae are briefly discussed.

\end{abstract}

\section{Introduction}

Kato et al. (2011) derived conditions of resonant excitation of disk oscillations in deformed disks
(see also Ferreira \& Ogilvie 2008 and Oktariani et al. 2008 for numerical studies and Kato 2011 for additional analytical studies).
The initial purpose of having examined this wave excitation process was to suggest a possible excitation mechanism of 
high-frequency quasi-periodic oscillations (QPOs) observed in black-hole low-mass X-ray binaries (BH LMXBs)
(see Kato 2004 and subsequent papers, for example, Kato 2008, 2012).
Not only our model but also many other models of HF QPOs in BH LMXBs have been proposed, but there is still no common consensus
on the origin of the QPOs.

To clarify whether the wave-wave resonant excitation process by Kato et al. (2011) is really one of realistic processes of 
excitation of disk oscillations in deformed disks,
it will be important to examine whether the mechanism can describe observations in systems where disks are definitely 
deformed and oscillations are observed.
A typical example of such objects is dwarf novae.
The gravitational fields of the disks of dwarf novae are tidally deformed by a secondary star, and if certain conditions are
satisfied the disks become unstable (tidal instability) and oscillatory phenomena (superhumps) appears.
Detailed observational studies on the oscillatory phenomena in dwarf novae have been performed by some groups, for example, by VSNET
group by T. Kato (a recent work by his group is, for example, Ohshima et al. 2012).

The tidal instability and superhump in dwarf navae are understood by the tidal instability (Lubow 1991) and precession of one-armed 
p-mode oscillation (Osaki 1985).
The wave-wave resonant excitation process can also describe them (Kato 2013).
In this contexts, we feel that the formulation by Kato et al. (2011) (hereafter paper I)
should be improved, since i) not all possible resonant cases are considered there, ii) in some parts, nonlinear treatment of
complex quantities was somewhat vague, and iii) interpretation of the instability was not enough.
The purpose of this paper is thus to refine the analyses in paper I and to make the meaning of the instability more clearer.
As an application, excitation of corrugation waves (warps) in dwarf novae disks, in addition to excitation of one-armed p-mode
oscillations, is briefly mentioned.

\section{Outline and Assumptions in This Paper}

On an axisymmetric steady disk a deformation is superposed.
The deformation is assumed to be time-periodic with frequency $\omega_{\rm D}$ and with azimuthal wavenumber $m_{\rm D}$.
In tidally deformed disks or in tilted disks, such deformation will be expected.
Even in other situations, long-living time-periodic deformations will be present on disks by a long-scale instability
of disks or by interactions with surrounding media.
Our basic assumption here is that such a time-periodic deformation is maintained on disks over a timescale 
longer than the growing or damping timescale of oscillatory perturbations on disks.
The displacement vector, $\mbox{\boldmath $\xi$}_{\rm D}(\mbox{\boldmath $r$}, t)$, associated with the deformation 
over the steady state is denoted by
\begin{equation}
    \mbox{\boldmath $\xi$}_{\rm D}(\mbox{\boldmath $r$}, t)=\Re\biggr[\hat{\mbox{\boldmath $\xi$}}_{\rm D}(\mbox{\boldmath $r$})
            {\rm exp}(i\omega_{\rm D}t)\biggr]
      =\Re\biggr[\breve{\mbox{\boldmath $\xi$}}_{\rm D}{\rm exp}[i(\omega_{\rm D}t-m_{\rm D}\varphi)]\biggr],
\label{2.1'}
\end{equation}
where $\Re$ denotes the real part, and $\varphi$ is the azimuthal coordinate of the cylindrical coordinates ($r$, $\varphi$, $z$) whose center 
is at the disk center and the $z$-axis is the rotating axis of the disk.

In addition to such a deformation, a set of two normal modes of oscillations are superposed on the axisymmetric steady disks.
The set of eigen-frequency and azimuthal wavenumber of these oscillations are denoted by
($\omega_1$, $m_1$) and ($\omega_2$, $m_2$).
The displacement vectors, $\mbox{\boldmath $\xi$}_{\rm i}(\mbox{\boldmath $r$}, t)$, associated with these 
oscillations are expressed as 
\begin{equation}
    \mbox{\boldmath $\xi$}_{\rm i}(\mbox{\boldmath $r$}, t)=\Re\biggr[\hat{\mbox{\boldmath $\xi$}}_{\rm i}
            (\mbox{\boldmath $r$}){\rm exp}(i\omega_{\rm i}t)\biggr]
      =\Re\biggr[\breve{\mbox{\boldmath $\xi$}}_{\rm i}{\rm exp}[i(\omega_{\rm i} t-m_{\rm i}\varphi)]\biggr] \qquad ({\rm i}=1, 2).
\label{2.2'}
\end{equation}

We now assume that the following resonant conditions among the two oscillations and the deformation are present:
\begin{equation}
    \omega_1+\omega_2+\omega_{\rm D}=\Delta\omega \quad {\rm and}\quad m_1+m_2+m_{\rm D}=0,
\label{2.3'}
\end{equation}
where $m_{\rm i}$'s (${\rm i}=1,2,$ and $D$) are integers.
In order to include in our formulation the cases where three frequencies are slightly deviated from the exact resonant
condition, $\Delta\omega$ is introduced in the first relation of equation (\ref{2.3'}), where $\vert\Delta\omega\vert$ 
is assumed to be much smaller than the absolute values of $\omega_1$ and $\omega_2$.\footnote{
The cases where $\omega_{\rm D}=0$ can be included in our analyses.
}
In the final analyses of instability, however, we consider only the case of $\Delta\omega=0$.
In order to represent the resonant conditions by a simple form without separately considering such cases as 
$\omega_2=\omega_1+\omega_{\rm D}$ and $\omega_2=-\omega_1+\omega_{\rm D}$, 
we have adopted such resonant forms as equations (\ref{2.3'}), where $\omega$'s and $m$'s  
can be taken to be positive or negative as long as the conditions (\ref{2.3'}) are satisfied.

In the non-linear stages, the $\omega_1$- and $\omega_2$-oscillations satisfying conditions (\ref{2.3'}) resonantly interact 
with each other through the disk deformation specified by $\omega_{\rm D}$.\footnote{
For oscillations to resonantly interact, additional conditions concerning wave forms of the oscillations in the vertical and 
radial directions are necessary.
These conditions are not mentioned here.
If these consitions are not satisfied, the value of the coupling term, $W$ or
$W^{\rm T}$, given by equation (\ref{3.3}) or (\ref{3.3'}) vanishes.  
}
Our main concern in  this paper is to examine 
how the amplitudes of the $\omega_1$- and
$\omega_2$-oscillations are changed by the quasi-nonlinear interactions with the disk deformation.
Quasi-nonlinear behavoirs of the $\omega_1$- and $\omega_2$-oscillations themselves are not considered, 
assuming that their amplitudes are smaller than that of the disk deformation.
It is further assumed that the amplitudes of the oscillations are so small that the amplitude of the disk deformation is unchanged 
by the resonant couplings with the oscillations.

In order to understand the results concerning the instability condition, however, 
it is instructive to consider the cases where the amplitude of 
the disk deformation is not kept unchanged, but changes as results of resonant interactions with the oscillations.
This is the case of three wave resonant interactions where no particular position is given to 
the disk deformation. 
This case is mentioned in section 5.
Application to excitation of superhumps in dwarf novae is briefly mentioned in section 6.     

\section{Formulation of Coupling Processes}
 
When we consider high-frequency QPOs in neutron-star and black-hole low-mass X-ray binaries, 
the effects of general relativity are important for quantitative studies.
The effects of general relativity, however, is not essential in understanding the essence of the instability
mechanism.
Hence, in this paper, for simplicity, we formulate the problem in the Newtonian frame.
We adopt a Lagrangian formulation introduced by Lynden-Bell and Ostriker (1987).

\subsection{Hydrodynamical Equations for Linear Oscillations and Disk Deformation}

The unperturbed disk with no deformation is steady and axisymmetric.
In the Lagrangian formulation, hydrodynamical perturbations, including a disk deformation, 
superposed over the unperturbed disks can be described by (Lynden-Bell \& Ostriker 1967)
\begin{equation}
       \frac{D_0^2\mbox{\boldmath $\xi$}}{Dt^2}=\delta\biggr(-\nabla\psi-\frac{1}{\rho}\nabla p\biggr),
\label{2.1}
\end{equation}
where $\mbox{\boldmath $\xi$}(\mbox{\boldmath $r$}, t)$ is a displacement vector associated with the perturbations, 
and $D_0/Dt$ is the time derivative along an unperturbed flow, $\mbox{\boldmath $u$}_0(\mbox{\boldmath $r$})$, and is related to 
the Eulerian time derivative, $\partial/\partial t$,  by
\begin{equation}
      \frac{D_0}{Dt}=\frac{\partial}{\partial t}+\mbox{\boldmath $u$}_0\cdot\nabla.
\label{2.2}
\end{equation}
In equation (\ref{2.1}) $\delta(X)$ represents the Lagrangian variation of $X$, and $\psi$ is the gravitational
potential.
Other notations in equation (\ref{2.1}) have their usual meanings.

In the cases of small amplitude, adiabatic and inviscid perturbations equation (\ref{2.1}) is written as (Lynden-Bell
\& Ostriker 1967) 
\begin{equation}
     \rho_0\frac{\partial^2\mbox{\boldmath $\xi$}}{\partial t^2}
       +2\rho_0(\mbox{\boldmath $u$}_0\cdot\nabla)\frac{\partial\mbox{\boldmath $\xi$}}{\partial t}
       +\mbox{\boldmath $L$}(\mbox{\boldmath $\xi$})=0,
\label{2.3}
\end{equation}
where $\mbox{\boldmath $L$}(\mbox{\boldmath $\xi$})$ is a linear Hermitian operator with respect to $\mbox{\boldmath $\xi$}$.
Equation (\ref{2.3}) is valid even when the perturbations are self-gravitating.
Hereafter, however, we consider only the cases of non-selfgravitating perturbations.
  
The disks where the behavoirs of oscillations are examined are not a steady equilibrium one, 
but in a deformed state as mentioned above.
If the disk deformation is maintained by a tidal force, for example, the deformation is a forced oscillation and the displacement vector, 
$\mbox{\boldmath $\xi$}_{\rm D}$, associated with the deformation is governed, in the linear approximations, by
\begin{equation}
     \rho_0\frac{\partial^2\mbox{\boldmath $\xi$}_{\rm D}}{\partial t^2}
       +2\rho_0(\mbox{\boldmath $u$}_0\cdot\nabla)\frac{\partial\mbox{\boldmath $\xi$}_{\rm D}}{\partial t}
       +\mbox{\boldmath $L$}(\mbox{\boldmath $\xi$}_{\rm D})=-\rho_0\nabla\psi_{\rm D},
\label{2.4}
\end{equation}
where $\psi_{\rm D}$ is the Eulerian variation of the gravitational potential, $\psi$,  due to the external tidal force, and
is expressed as
\begin{equation}
     \psi_{\rm D}(\mbox{\boldmath $r$},t)=\Re\biggr[\hat{\psi}_{\rm D}(\mbox{\boldmath $r$})
            {\rm exp}(i\omega_{\rm D}t)\biggr]
            =\Re\biggr[\breve{\psi}_{\rm D}{\rm exp}[i(\omega_{\rm D}t-m_{\rm D}\varphi)]\biggr],
\label{2.4'}
\end{equation}
where $\omega_{\rm D}$ is the tidal frequency associated with azimuthal wavenumber $m_{\rm D}$, and
$\omega_{\rm D}$ and $m_{\rm D}$ are taken to be related by $\omega_{\rm D}=m_{\rm D}\Omega_{\rm orb}$, 
$\Omega_{\rm orb}$ being the angular velocity of rotation of the secondary star.

If no external force acts on the disk,   
the displacement vector, $\mbox{\boldmath $\xi$}_{\rm D}$, associated with the deformation is 
governed by an equation having the same form as equation (\ref{2.3}) in the linear approximation.

\subsection{Quasi-Nonlinear Coupling among Oscillations and Disk Deformation}

Now, we consider two normal modes of oscillations, $\mbox{\boldmath $\xi$}_1$ and $\mbox{\boldmath $\xi$}_2$,
with ($\omega_1$, $m_1$) and ($\omega_2$, $m_2$).
In the linear stage, the perturbation, $\mbox{\boldmath $\xi$}$, imposed on a deformed disk is simply the sum 
of these two oscillations:
\begin{eqnarray}
        \mbox{\boldmath $\xi$}_1(\mbox{\boldmath $r$}, t) 
          = &&A_1 \mbox{\boldmath $\xi$}_1(\mbox{\boldmath $r$}, t)
            + A_2 \mbox{\boldmath $\xi$}_2(\mbox{\boldmath $r$}, t)      
                                  \nonumber    \\
          =&&\Re\biggr[A_1\breve{\mbox{\boldmath $\xi$}}_1{\rm exp}[i(\omega_1t-m_1\varphi)]
                      +A_2\breve{\mbox{\boldmath $\xi$}}_2{\rm exp}[i(\omega_2t-m_2\varphi)]\biggr],
\label{2.6}
\end{eqnarray}
where $A_1$ and $A_2$ are amplitudes and are arbitrary constants at the linear stage.
In the quasi-nonlinear stage, the oscillations  are modified through quasi-nonlinear couplings with disk deformation, 
$\mbox{\boldmath $\xi$}_{\rm D}$, so that they satisfy a quasi-nonlinear wave equation. 
In the case where the deformation is internally maintained (i.e., no external force), the non-linear wave equation is
\begin{equation}
    \rho_0\frac{\partial^2\mbox{\boldmath $\xi$}}{\partial t^2}
       +2\rho_0(\mbox{\boldmath $u$}_0\cdot\nabla)\frac{\partial\mbox{\boldmath $\xi$}}{\partial t}
       +\mbox{\boldmath $L$}(\mbox{\boldmath $\xi$})
    = \mbox{\boldmath $C$}(\mbox{\boldmath $\xi$}, \mbox{\boldmath $\xi$}_{\rm D}),
\label{2.7}
\end{equation}
where $\mbox{\boldmath $C$}$ is the quasi-nonlinear coupling terms and is given by (Kato 2004, 2008)
\begin{eqnarray}
     C_i(\mbox{\boldmath $\xi$}, \mbox{\boldmath $\xi$}_{\rm D})=&&-\rho_0\xi_j\xi_{{\rm D}k}\frac{\partial^3}
         {\partial r_i\partial r_j\partial r_k}\psi_0
         -\frac{\partial}{\partial r_k}\biggr(p_0\frac{\partial\xi_k}{\partial r_j}
          \frac{\partial\xi_{{\rm D}j}}{\partial r_i}
          +p_0\frac{\partial\xi_{{\rm D}k}}{\partial r_j}\frac{\partial\xi_j}{\partial r_i}\biggr)    \nonumber \\
      &&+\frac{\partial}{\partial r_j}\biggr[(\Gamma_1-1)p_0\biggr(\frac{\partial \xi_j}{\partial r_i}{\rm div}\mbox{\boldmath $\xi$}_{\rm D}
                +\frac{\partial\xi_{{\rm D}j}}{\partial r_i}{\rm div}\mbox{\boldmath $\xi$}\biggr)\biggr]
                        \nonumber   \\
      &&+\frac{\partial}{\partial r_i}\biggr[(\Gamma_1-1)p_0\frac{\partial\xi_k}{\partial r_j}\frac{\partial\xi_{{\rm D}j}}{\partial r_k}\biggr]
        +\frac{\partial}{\partial r_i}\biggr[\Gamma_1(\Gamma_1-1)
                    p_0{\rm div}\mbox{\boldmath $\xi$}\cdot{\rm div}\mbox{\boldmath $\xi$}_{\rm D}\biggr],
\label{2.8}
\end{eqnarray}
where the subscript $i$ attached to $C$ and $r$ denotes the $i$-component of vectors $\mbox{\boldmath $C$}$ and $\mbox{\boldmath $r$}$
in the Cartesian coordinates, and we use Einstein's convention of indices (i.e., take summaation if a term has the same index 
variable twice).
Furthermore, $\Gamma_1$ is the barotropic index specifying
the linear part of the relation between the Lagrangian variations of pressure, $\delta p$, and density, $\delta\rho$,
i.e., $\delta p/p_0=\Gamma_1\delta\rho/\rho_0$.
 
In the cases where the disk deformation is due to an external force, additional terms should be added to equation 
(\ref{2.8}). 
In the case of tidal deformation, for example, the Eulerian variation of the gravitational potential, $\psi_{\rm D}$, 
is associated with $\mbox{\boldmath $\xi$}_{\rm D}$.
Because of this, in considering the Lagrangian variation of $\nabla\psi$ [see $\delta\nabla\psi$ in equation (\ref{2.1})], 
a quasi-nonlinear term of $\xi_j\partial \nabla\psi_{\rm D}/\partial r_j$, which has been neglected in deriving equation (\ref{2.8}),
should be taken into account, since we have
\begin{equation}
       \delta\nabla\psi=\nabla\psi_{\rm D}+\xi_j\frac{\partial}{\partial r_j}\nabla(\psi_0+\psi_{\rm D})
           +\frac{1}{2}\xi_j\xi_k\frac{\partial^2}{\partial r_j\partial r_k}\nabla(\psi_0+\psi_{\rm D})+...,
\label{2.9}
\end{equation}
where $\psi_0$ is the unperturbed gravitational potential in the unperturbed axisymmetric disk.
Corresponding to this, the coupling term $\mbox{\boldmath $C$}$ is changed to $\mbox{\boldmath $C$}^{\rm T}$, which is
\begin{equation}
       \mbox{\boldmath $C$}^{\rm T}(\mbox{\boldmath $\xi$}, \mbox{\boldmath $\xi$}_{\rm D})
         =\mbox{\boldmath $C$}(\mbox{\boldmath $\xi$}, \mbox{\boldmath $\xi$}_{\rm D})-\rho_0\xi_j\frac{\partial}{\partial r_j}\nabla\psi_{\rm D}.
\label{2.10}
\end{equation}
Here and hereafter, the superscript ${\rm T}$ is attached to $\mbox{\boldmath $C$}$, when the disk deformation is due to  
tidal force, and the quasi-nonlinear wave equation is 
\begin{equation}
    \rho_0\frac{\partial^2\mbox{\boldmath $\xi$}}{\partial t^2}
       +2\rho_0(\mbox{\boldmath $u$}_0\cdot\nabla)\frac{\partial\mbox{\boldmath $\xi$}}{\partial t}
       +\mbox{\boldmath $L$}(\mbox{\boldmath $\xi$})
    = \mbox{\boldmath $C$}^{\rm T}(\mbox{\boldmath $\xi$}, \mbox{\boldmath $\xi$}_{\rm D}).
\label{2.7'}
\end{equation}

The disk oscillations, $\mbox{\boldmath $\xi$}(\mbox{\boldmath $r$}, t)$, resulting from the quasi-nonlinear coupling 
through disk deformation, $\mbox{\boldmath $\xi$}_{\rm D}(\mbox{\boldmath $r$}, t)$, will be written generally in the form:\footnote{
The expression for $\mbox{\boldmath $\xi$}(\mbox{\boldmath $r$},t)$ in Kato et al. (2011) was somewhat vague.
}
\begin{eqnarray}
      \mbox{\boldmath $\xi$}(\mbox{\boldmath $r$},t)=&&\Re \sum_{{\rm i}=1}^2}A_{\rm i}(t)\hat{\mbox{\boldmath $\xi$}}_{\rm i}(\mbox{\boldmath $r$})
                       {\rm exp(i\omega_{\rm i} t)    
               +\Re \sum_i^{2}\sum_{\alpha\not= 1,2} A_{i,\alpha}\hat{\mbox{\boldmath $\xi$}}_\alpha(\mbox{\boldmath $r$}){\rm exp}(i\omega_{\rm i} t)
                             \nonumber   \\
             &&+ {\rm oscillating \ terms\ with \ other \ frequencies}.
\label{2.11}
\end{eqnarray}
The original two oscillations, $\mbox{\boldmath $\xi$}_1$ and $\mbox{\boldmath $\xi$}_2$, resonantly interact through the disk deformation.
Hence, their amplitudes secularly change with time, which is taken into account in equation (\ref{2.11}) by taking
the amplitudes, $A_{\rm i}$'s, to be slowly varying functions of time.
The spatial forms of the terms whose time-dependence is exp($i\omega_{\rm i} t$) but whose spatial dependence is different from 
$\hat{\mbox{\boldmath $\xi$}}_{\rm i}$ are expressed by a sum of a series of eigen-functions, $\hat{\mbox{\boldmath $\xi$}}_\alpha$
 ($\alpha\not= 1$ and 2),
assuming that they make a complete set.
The terms  whose time-dependences are different from exp$(i\omega_{\rm i}t)$ are not written down explicitly in equation (\ref{2.11}),
since these terms disappear by taking long-term time average when we are interested in phenomena with frequencies 
of $\omega_{\rm i}$ (i $=$ 1 and 2).

In order to derive equations describing the time evolution of $A_{\rm i}(t)$, we substitute equation (\ref{2.11}) into the
left-hand side of equation (\ref{2.7}) or (\ref{2.7'}).
Then, considering that $\hat{\mbox{\boldmath $\xi$}}_{\rm i}$'s and $\hat{\mbox{\boldmath $\xi$}}_\alpha$'s are displacement vectors 
associated with eigen-functions of the linear wave equation (\ref{2.3}), we find that the left-hand side of equation (\ref{2.7}) 
or (\ref{2.7'}) becomes the real part of 
\begin{eqnarray}
   &&2\rho_0\sum_{i=1}^2\frac{dA_{\rm i}}{dt}\biggr[i\omega_{\rm i}+(\mbox{\boldmath $u$}_0\cdot\nabla)\biggr]
                     \hat{\mbox{\boldmath $\xi$}}_{\rm i}{\rm exp}(i\omega_{\rm i}t)
                    \nonumber       \\
   &&+\rho_0\sum_i\sum_\alpha A_{i,\alpha}\biggr[(\omega_\alpha^2-\omega_{\rm i}^2)-2i(\omega_\alpha-\omega_{\rm i})(\mbox{\boldmath $u$}_0\cdot\nabla)\biggr]
         \hat{\mbox{\boldmath $\xi$}}_\alpha{\rm exp}(i\omega_{\rm i}t),
\label{2.12}
\end{eqnarray}
where $d^2A_{\rm i}/d t^2$ has been neglected, since $A_{\rm i}(t)$($i=1$ and 2) are slowly varying functions of time.
Now, the real part of equation (\ref{2.12}) is integrated over the whole volume of disks after being multiplied by 
$\mbox{\boldmath $\xi$}_1[=\Re\ \hat{\mbox{\boldmath $\xi$}}_1{\rm exp}(i\omega_1t)]$.\footnote{
The formula
$$
      \Re(A)\Re(B)=\frac{1}{2}\Re[AB+AB^*]=\frac{1}{2}\Re[AB+A^*B]      \nonumber
$$
is used, where $A$ and $B$ are complex variables and $B^*$ is the complex conjugate of $B$.
}
Then, the term resulting from the second term of equation (\ref{2.12}) vanishes [see an orthogonal relation given by
equation (8) of paper I] and the results become
\begin{equation}
      \Re \ i\frac{dA_1}{dt}\biggr\langle \rho_0\hat{\mbox{\boldmath $\xi$}}_1^*[\omega_1-i(\mbox{\boldmath $u$}_0\cdot\nabla)]
                \hat{\mbox{\boldmath $\xi$}}_1\biggr\rangle,
\label{2.13}
\end{equation}
where $\langle X \rangle$ denotes the volume integration of $X(\mbox{\boldmath $r$})$, and
the asterisk * denotes the complex conjugate.
This can be further reduced to
\begin{equation}
    \Re\  i\frac{2E_1}{\omega_1}\frac{dA_1}{dt}
\label{}
\end{equation}
by using the wave energy of the oscillation of $\mbox{\boldmath $\xi$}_1$, $E_1$, which is a real quantity defined by (e.g., Kato 2001)
\begin{equation}
      E_1=\frac{1}{2}\omega_1\biggr[\omega_1\langle\rho_0\hat{\mbox{\boldmath $\xi$}}_1^*\hat{\mbox{\boldmath $\xi$}}_1\rangle
          -i\langle\rho_0\hat{\mbox{\boldmath $\xi$}}_1^*(\mbox{\boldmath $u$}_0\cdot\nabla)\hat{\mbox{\boldmath $\xi$}}_1\rangle\biggr].
\label{2.14}
\end{equation}
It is of importance to note that the wave energy can be expressed in the case of the oscillations in geometrically thin disks
as (e.g., see Kato 2001)
\begin{equation}
    E_1= \frac{\omega_1}{2}\biggr\langle(\omega_1-m_1\Omega)\rho_0(\hat{\xi}_{1,r}^*\hat{\xi}_{1,r}
              +\hat{\xi}_{1,z}^*\hat{\xi}_{1,z})\biggr\rangle,
\label{energy}
\end{equation}
where $\Omega(r)$ is the angular velocity of disk rotation.

Now, the real part of equation (\ref{2.7}) is multiplied by  
$\mbox{\boldmath $\xi$}_1(\mbox{\boldmath $r$})$ and integrated over the whole volume.
Then, considering the results of the above paragraph, we have
\begin{equation}
     \Re \ i\frac{2E_1}{\omega_1}\frac{dA_1}{dt}
      =\frac{1}{2}\Re \ \biggr[A_2(t)A_{\rm D}\biggr\langle\hat{\mbox{\boldmath $\xi$}}_1\cdot
         \mbox{\boldmath $C$}(\hat{\mbox{\boldmath $\xi$}}_2,\hat{\mbox{\boldmath $\xi$}}_{\rm D})\biggr\rangle
            {\rm exp}(i\Delta \omega t)\biggr],
\label{2.15}
\end{equation}
where $\mbox{\boldmath $C$}(\hat{\mbox{\boldmath $\xi$}}_2,\hat{\mbox{\boldmath $\xi$}}_{\rm D})$ is replaced to 
$\mbox{\boldmath $C$}^{\rm T}(\hat{\mbox{\boldmath $\xi$}}_2,\hat{\mbox{\boldmath $\xi$}}_{\rm D})$, when tidal deformation 
is considered.
In deriving the right-hand side of equation (\ref{2.15}), the time periodic terms with high frequencies (i.e., non-resonant terms) have
been neglected by time average being taken.

Similarly, we multiply $\mbox{\boldmath $\xi$}_2(\mbox{\boldmath $r$})$ to the real part of equation (\ref{2.7}) and integrate over the whole volume
to lead to
\begin{equation}
     \Re \ i\frac{2E_2}{\omega_2}\frac{dA_2}{dt}
      =\frac{1}{2}\Re \ \biggr[A_1(t)A_{\rm D}\biggr\langle\hat{\mbox{\boldmath $\xi$}}_2\cdot \mbox{\boldmath $C$}(\hat{\mbox{\boldmath $\xi$}}_1,
           \hat{\mbox{\boldmath $\xi$}}_{\rm D})\biggr\rangle
            {\rm exp}(i\Delta \omega t)\biggr].
\label{2.16}
\end{equation}
As mentioned above, in the case of tidal deformation, $\mbox{\boldmath $C$}(\hat{\mbox{\boldmath $\xi$}}_1,\hat{\mbox{\boldmath $\xi$}}_{\rm D})$
is replaced by $\mbox{\boldmath $C$}^{\rm T}(\hat{\mbox{\boldmath $\xi$}}_1,\hat{\mbox{\boldmath $\xi$}}_{\rm D})$.

\section{Growth of Resonant Oscillations}

Since the amplitude $A_{\rm D}$ is assumed to be constant, the time evolutions of $A_1$ and $A_2$ are determined by solving the
simultaneous equations (\ref{2.15}) and (\ref{2.16}).
What are governed by equations (\ref{2.15}) and (\ref{2.16}) are the imaginary part of $A_1$ and $A_2$, i.e., $A_{1{\rm i}}$ 
and $A_{2{\rm i}}$.
Their real parts are not related to the resonance, and we can neglect them in considering the exponential growth (or damping)
of $A_{1{\rm i}}$ and $A_{2{\rm i}}$.
Hereafter, we restrict our attention only to the case of $\Delta\omega=0$.
Then, we have from equations (\ref{2.15}) and (\ref{2.16})
\begin{equation}
     -\frac{2E_1}{\omega_1}\frac{A_{1,{\rm i}}}{dt}=-\frac{1}{2}A_{2,{\rm i}}\Im (A_{\rm D}W),
\label{3.1}
\end{equation}
\begin{equation}
     -\frac{2E_2}{\omega_2}\frac{A_{2,{\rm i}}}{dt}=-\frac{1}{2}A_{1,{\rm i}}\Im (A_{\rm D}W),
\label{3.2}
\end{equation}
where 
\begin{equation}
      W\equiv\biggr\langle\hat{\mbox {\boldmath $\xi$}}_1\cdot\mbox {\boldmath $C$}
                         (\hat{\mbox {\boldmath $\xi$}}_2, \hat{\mbox {\boldmath $\xi$}}_{\rm D})\biggr\rangle
            =\biggr\langle\hat{\mbox {\boldmath $\xi$}}_2\cdot\mbox {\boldmath $C$}
                         (\hat{\mbox {\boldmath $\xi$}}_1, \hat{\mbox {\boldmath $\xi$}}_{\rm D})\biggr\rangle.
\label{3.3}
\end{equation}
The last equality showing commutability of $\hat{\mbox{\boldmath $\xi$}}_1$ and $\hat{\mbox{\boldmath $\xi$}}_2$
is important for obtaining an expression for instability.
This commutative relation is derived by using an expression for $\mbox{\boldmath $C$}$
given by equation (\ref{2.8}) and performing the volume integration by part, neglecting surface integrals under the assumption that
the unperturbed density and pressure vanish on the surface of the system.
In the case where the deformation is due to the tidal force, $W$'s in equations (\ref{3.1}) and (\ref{3.2}) 
are replaced by $W^{\rm T}$ which is 
\begin{equation}
      W^{\rm T}\equiv\biggr\langle\hat{\mbox {\boldmath $\xi$}}_1\cdot\mbox {\boldmath $C$}^{\rm T}
             (\hat{\mbox {\boldmath $\xi$}}_2, \hat{\mbox {\boldmath $\xi$}}_{\rm D})\biggr\rangle
      =\biggr\langle\hat{\mbox {\boldmath $\xi$}}_2\cdot\mbox {\boldmath $C$}^{\rm T}
             (\hat{\mbox {\boldmath $\xi$}}_1, \hat{\mbox {\boldmath $\xi$}}_{\rm D})\biggr\rangle.
\label{3.3'}
\end{equation}
The commutability in equation (\ref{3.3'}) is obtained by using $\mbox{\boldmath $C$}^{\rm T}$ given by equation (\ref{2.10}).

Eliminating $A_{2,{\rm i}}$ from equations (\ref{3.1}) and (\ref{3.2}), we have
\begin{equation}
     \frac{d^2A_{1,{\rm i}}}{dt^2}
          =\frac{1}{16}\biggr(\frac{E_1E_2}{\omega_1\omega_2}\biggr)^{-1}\biggr[\Im (A_{\rm D}W)\biggr]^2A_{1,{\rm i}}.
\label{3.4}
\end{equation}
The same equation is obtained for $A_{2,{\rm i}}$ by eliminating $A_{1,{\rm i}}$ instead of $A_{2,{\rm i}}$.

Equation (\ref{3.4}) shows that if 
\begin{equation}
       \frac{E_1E_2}{\omega_1\omega_2}>0,
\label{3.5}
\end{equation}
the oscillations grow with the growth rate given by
\begin{equation}
      \biggr(\frac{\omega_1\omega_2}{16E_1E_2}\biggr)^{1/2}\vert\Im(A_{\rm D}W)\vert.
\label{3.6}
\end{equation}
In the case where the disk deformation is due to tidal force, $W$ in equation (\ref{3.6}) is changed to $W^{\rm T}$.

At a glance, the sign of the instability condition given by equation (\ref{3.5}) is opposite to that in paper I.
This is due to a formal difference of expression for the resonant condition, and there is no difference in contents.
In paper I we have examined two resonant cases of $\omega_2^{\rm I}=\omega_1^{\rm I}+\omega_{\rm D}^{\rm I}$ and 
$\omega_2^{\rm I}=\omega_1^{\rm I}-\omega_{\rm D}^{\rm I}$ (the superscript I is attached to 
$\omega$'s in order to distinct these $\omega$'s from those in the resonant condition of $\omega_1+\omega_2+\omega_{\rm D}=0$ 
in the present paper),
and derived the instability condition: $E_1E_2/\omega_1^{\rm I}\omega_2^{\rm I}<0$. 
The resonant case of $\omega_2^{\rm I}=\omega_1^{\rm I}+\omega_{\rm D}^{\rm I}$ in paper I 
corresponds to the case where $\omega_2$ in the present paper is formally written as $-\omega_2^{\rm I}$.
If this change is adopted the present instability condition becomes $E_1E_2/\omega_1\omega_2^{\rm I}<0$, and is the same as that in paper I.
Furthermore, the case of $\omega_2^{\rm I}=\omega_1^{\rm I}-\omega_{\rm D}^{\rm I}$ in paper I corresponds to the case where
$\omega_1$ in the present paper is written as $-\omega_1^{\rm I}$ and we have $E_1E_2/\omega_1^{\rm I}\omega_2<0$.

In principle, the cases of $\omega_2^{\rm I}=-(\omega_1^{\rm I}+\omega_{\rm D}^{\rm I})$ and 
$\omega_2^{\rm I}=-(\omega_1^{\rm I}-\omega_{\rm D}^{\rm I})$
are possible, but not examined in paper I.
If we consider these cases by the same way as in paper I, we obtain $E_1E_2/\omega_1^{\rm I}\omega_2^{\rm I}>0$,
different from the cases of $\omega_2^{\rm I}=\omega_1^{\rm I}+\omega_{\rm D}^{\rm I}$ and 
$\omega_2^{\rm I}=\omega_1^{\rm I}-\omega_{\rm D}^{\rm I}$.
This instability condition is also consistent with the condition, $E_1E_2/\omega_1\omega_2>0$, of the present paper.
This is because in these cases $\omega_1=\omega_1^{\rm I}$ and $\omega_2=\omega_2^{\rm I}$ 
(the sign of $\omega_{\rm D}$ does not appear in the instability criterion).
In summary, in all resonant cases the instability condition is summaried by equation (\ref{3.5}), if the resonant condition is 
written as $\omega_1+\omega_2+\omega_{\rm D}=0$.

\section{Three Wave Interactions and Instability}

In the case where $\vert\omega_{\rm D}\vert$ is small, the resonance occurs between two oscillations with opposite signs of $\omega$'s,
i.e., $\omega_1\omega_2<0$. 
In this case the instability condition (\ref{3.5}) is written as 
\begin{equation}
      E_1E_2<0.
\label{D.4}
\end{equation}
This might suggest that the instability is, at the lowest order of approximation, a result of energy exchange between two oscillations 
through a catalytic action of disk deformation.
Positive energy flows from a negative-energy oscillation to a positive-energy oscillation, leading to growth of both oscillations.
If $\omega_1>0$ and $\omega_2<0$ (and also $m_1>0$ and $m_2<0$), i.e., if both oscillations are prograde with opposite signs of frequency, 
we introduce  $\tilde {\omega}_2$ and $\tilde{m}_2$ defined,
respectively, by $-\omega_2$ and $-m_2$ so that all $\omega$'s and $m$'s  become positive.
Then, the instability condition (\ref{D.4}) is reduced to
\begin{equation}
     \langle\omega_1-m_1\Omega\rangle \langle\tilde{\omega}_2-\tilde{m}_2\Omega\rangle < 0,
\label{D.5}
\end{equation}
where $\langle\ X\ \rangle$ is a weighted mean of $X$ in the region where oscillations prevails [see expression (\ref{energy}) for 
wave energy].
This equation shows that for the instability to occur, one oscillation is inside of its corotation radius, 
while the other must be outside of its corotation radius.\footnote{
The corotation radius of an oscillation with $\omega$ and $m$ is defined as the radius where $\omega=m\Omega$.
Inside the radius $\omega-m\Omega<0$, while outside the radius $\omega-m\Omega>0$.
}

The cause of instability is, however, not always interpreted in terms of energy exchange between $\omega_1$- and $\omega_2$-oscillations
alone.
Energy exchange between deformation and oscillation is of importance
in the case where $\vert\omega_{\rm D}\vert$ is large compared with $\vert\omega_1\vert$ and $\vert\omega_2\vert$.\footnote{
The tidal instability in dwarf novae corresponds to this case.
In paper I this case was not explicitly considered.
}
In this case the resonance ($\omega_1+\omega_2+\omega_{\rm D}=0$) occurs for oscillations with the same signs of $\omega_1$ and $\omega_2$,
and the instability condition (\ref{3.5}) is written as
\begin{equation}
      E_1E_2>0,
\label{D.6}
\end{equation}
or
\begin{equation}
      \langle\omega_1-m_1\Omega\rangle \langle\omega_2-m_2\Omega\rangle > 0,
\label{D.7}
\end{equation}
distinct from the case where $\vert\omega_{\rm D}\vert$ is small.
Equation (\ref{D.6}) clearly shows that in the present case of $\vert\omega_{\rm D}\vert$ being large, the instability
cannot be interpreted as energy exchange between two oscillations.
That is, both oscillations grow by getting energy from the deformation (in the case of $E_1>0$ and $E_2>0$), or
by lossing energy to the deformation (in the case of $E_1<0$ and $E_2<0$).
In other words, energy exchange between oscillations and deformation is essential for growth of the oscillations. 

Unless the disk deformation is maintained like the case of tidal deformation, or the deformation has a
sufficiently large amplitude,
the effects of amplitude change of the disk deformation on oscillations will be non-negligible.
To examine this, let us consider the case where the deformation is not maintained, but simply follows the same 
hydrodymnamical equation as those of the oscillations.
Then, by similar processes with those used to derive equation (\ref{2.15}) and (\ref{2.16}), we have
\begin{equation}
     \Re i\frac{2E_{\rm D}}{\omega_{\rm D}}\frac{dA_{\rm D}}{dt}=
          \frac{1}{2}\Re\ \biggr[A_1(t)A_2(t)
     \biggr\langle\hat{\mbox {\boldmath $\xi$}}_{\rm D}\cdot\mbox {\boldmath $C$}
             (\hat{\mbox {\boldmath $\xi$}}_1, \hat{\mbox {\boldmath $\xi$}}_2)\biggr\rangle{\rm exp}(i\Delta\omega t)\biggr],
\label{D.7'}
\end{equation}
where $E_{\rm D}$ is the wave energy of disk deformation.
In this case we have the commutative relation:
\begin{equation} 
            \biggr\langle\hat{\mbox {\boldmath $\xi$}}_{\rm D}\cdot\mbox {\boldmath $C$}
             (\hat{\mbox {\boldmath $\xi$}}_1, \hat{\mbox {\boldmath $\xi$}}_2)\biggr\rangle
          =\biggr\langle\hat{\mbox {\boldmath $\xi$}}_1\cdot\mbox {\boldmath $C$}
             (\hat{\mbox {\boldmath $\xi$}}_2, \hat{\mbox {\boldmath $\xi$}}_{\rm D})\biggr\rangle
          =\biggr\langle\hat{\mbox {\boldmath $\xi$}}_2\cdot\mbox {\boldmath $C$}
             (\hat{\mbox {\boldmath $\xi$}}_1, \hat{\mbox {\boldmath $\xi$}}_{\rm D})\biggr\rangle
                                     \equiv W.
\label{D.8}
\end{equation}
Hence, if $\Delta\omega=0$,  we have the following set of equations [see equations (\ref{3.1}) and (\ref{3.2})]:
\begin{eqnarray}
    && -\frac{2E_1}{\omega_1}\frac{dA_{1,{\rm i}}}{dt}=-\frac{1}{2}A_{2,{\rm i}}A_{D,{\rm i}}\Re W,   \label{D.9}  \\    
    && -\frac{2E_2}{\omega_2}\frac{dA_{2,{\rm i}}}{dt}=-\frac{1}{2}A_{1,{\rm i}}A_{D,{\rm i}}\Re W,   \label{D.10} \\      
    && -\frac{2E_D}{\omega_D}\frac{dA_{D,{\rm i}}}{dt}=-\frac{1}{2}A_{1,{\rm i}}A_{2,{\rm i}}\Re W.   \label{D.11}
\end{eqnarray}

In order to examine how the amplitude, $A_{D,{\rm i}}(t)$, changes with time, we take the time derivative of equation (\ref{D.11})
and substitute equations (\ref{D.9}) and (\ref{D.10}) into the resulting equation.
Then we have
\begin{equation}
      \frac{d^2A_{D, {\rm i}}}{dt^2}
       = \frac{1}{16}\biggr(\frac{E_2E_D}{\omega_2\omega_D}\biggr)^{-1}\biggr[A_{1,{\rm i}}^2
            +\frac{E_2/\omega_{2}}{E_1/\omega_{1}}A_{2,{\rm i}}^2\biggr]
           (\Re W)^2 A_{D,{\rm i}}.
\label{D.12}
\end{equation}
This equation shows that in the case of $(E_1/\omega_1)(E_2/\omega_2)>0$, $A_{D,{\rm i}}$ damps with time unless
$(E_D/\omega_D)(E_2/\omega_2)>0$.
However, $(E_1/\omega_1)(E_2/\omega_2)>0$ and $(E_D/\omega_D)(E_2/\omega_2)>0$ cannot be realized usually at the same time 
in resonance.
This comes from the following situation.
The sign of $E_D/\omega_D$ is equal to the sign of $\langle \omega_D-m_D\Omega \rangle$, where
$\langle \ X \ \rangle$ is a some spatial average of $X$ [remember the definintion of wave energy, i.e., see
equation ({\ref{energy})].
Hence, when the resonant conditions are satisfied, we have 
\begin{equation}
    {\rm Sign}\ \langle\omega_D-m_D\Omega\rangle = -{\rm Sign}\ [\langle\omega_1-m_1\Omega\rangle+\langle\omega_2-m_2\Omega\rangle]. 
\label{D.13}
\end{equation}
Roughly speaking, the signs of $\langle\omega_1-m_1\Omega\rangle$ and $\langle\omega_2-m_2\Omega\rangle$ are equal,
respectively, to the signs of $E_1/\omega_1$ and $E_2/\omega_2$ [see equation (\ref{energy})].
This means 
\begin{eqnarray}
    &&{\rm Sign}\ \frac{E_D}{\omega_D} < 0, \quad\quad {\rm if}\ \frac{E_1}{\omega_1}>0 \ {\rm and}\ \frac{E_2}{\omega_2}>0, \\
    &&{\rm Sign}\ \frac{E_D}{\omega_D} > 0, \quad\quad {\rm if}\ \frac{E_1}{\omega_1}<0 \ {\rm and}\ \frac{E_2}{\omega_2}<0.
\end{eqnarray}
In both cases we have $(E_D/\omega_D)(E_2/\omega_2)<0$.
In other words, when $\omega_1$- and $\omega_2$-oscillations grow by $(E_1/\omega_1)(E_2/\omega_2)>0$, $\omega_D$-oscillation
damps.

This situation can be also demonstrated clearly from the following considerations.
Equations (\ref{D.9}) -- (\ref{D.11}) are a set of simultaneous non-linear differential equations with respect to $A_{1,{\rm i}}$,
$A_{2,{\rm i}}$, and $A_{{\rm D},{\rm i}}$.
By multiplying $A_{1,{\rm i}}$ and $A_{2,{\rm i}}$, respectively, to equation (\ref{D.9}) and (\ref{D.10}), and
taking the difference of the resulting two equations, we have
\begin{equation}
    \frac{E_1}{\omega_1}A_{1,{\rm i}}^2-\frac{E_2}{\omega_2}A_{2,{\rm i}}^2= \ {\rm const.}   
\label{D.14}  
\end{equation}
By the similar procedures, we also have
\begin{equation}
      \frac{E_1}{\omega_1}A_{1,{\rm i}}^2-\frac{E_{\rm D}}{\omega_{\rm D}}A_{{\rm D},{\rm i}}^2= \ {\rm const.}   
\label{D.15} 
\end{equation}
\begin{equation}
      \frac{E_2}{\omega_2}A_{2,{\rm i}}^2-\frac{E_{\rm D}}{\omega_{\rm D}}A_{{\rm D},{\rm i}}^2= \ {\rm const.},  
\label{D.16}
\end{equation}
where the right-hand sides of equations (\ref{D.14}) -- (\ref{D.16}) are time-independent constants.
Equation (\ref{D.14}) shows that when $(E_1/\omega_1)(E_2/\omega_2)>0$, $A_{1,{\rm i}}$ and $A_{2,{\rm i}}$ can increase with time.
Equation (\ref{D.15}), however, demonstrates that in that case $A_{D,{\rm i}}$ decreases since the sign of $E_2/\omega_2$ and
$E_D/\omega_D$ is opposite.

More generally speaking, let us consider resonant interactions among three oscillations.
Then, two of them can grow by consumption of the other oscillation, but not all oscillations can grow.

\section{Applications to Superhumps in Dwarf Novae}

Typical objects with deformed disks which are maintained are dwarf novae, where the gravitational potential in disks is 
tidally deformed by a secondary star.
Whitehurst (1988) found that under a certain condition the disks become unstable to an eccentric deformation having slow
precession.
The instability is known to be due to a tidal instability of the so-called 3 : 1 resonance (Lubow 1991). 
The precession comes from excitation of one-armed oscillation and it brings about superhump phenomena (Osaki 1985).
(See Osaki 1996 for review of superoutburst and superhump of dwarf novae.)
Lubow (1992) further shows that a tilt is also excited at the 3 : 1 resonance by a mode-coupling process similar
to that in the case of excitation of the eccentric deformation.
The present wave-wave resonant instability process can also describe the instability and 
excitations of both one-armed oscillation and tilt at the 3 : 1 resonance.
Although the excitation of one-armed oscillation was already discussed in Kato (2013), we briefly 
summarize here the essence of excitation of both modes.  

\subsection{Tidal Instability and Excitation of Low-Frequency One-Armed p-Mode Oscillation}

In tidally deformed disks, $m_{\rm D}$ and $\omega_{\rm D}$ are related by $\omega_{\rm D}=m_{\rm D}\Omega_{\rm orb}$, when
the orbit of the secondary star is circular, where
$\Omega_{\rm orb}$ is the orbital frequency of the secondary star and $m_{\rm D}$ is an integer.
As the $\omega_1$-oscillation, we take one-armed ($m_1=1$) low-frequency p-mode oscillation with a positive frequency.
The dispersion relation\footnote{
Let us consider vertically isothermal disks.
Then, the dispersion relation for perturbations which are local in the radial direction is given by
$$
    [(\omega-m\Omega)^2-\kappa^2][(\omega-m\Omega)^2-n\Omega_\bot^2]=k^2c_{\rm s}^2(\omega-m\Omega)^2,
$$
where $\Omega_\bot$ is the vertical epicyclic frequency, $n$ an integer representing the node number of oscillations
in the vertical direction, $k$ the radial wavenumber, and $c_{\rm s}$ the acoustic speed in disks
(see, e.g., Kato 2001 or Kato et al. 2008).
It is noticed that $\Omega_\bot>\Omega > \kappa$ in geometrically thin disks.
}
shows that the frequency, $\omega_1$, must be smaller than $\Omega-\kappa$, i.e., $\omega_1<\Omega-\kappa$,  
where $\Omega(r)$ is the angular velocity of disk rotation and $\kappa(r)$ is the epicyclic frequency in the equatorial plane.
Since $\Omega(r)-\kappa(r)$ is positive and increases outwards in tidally deformed disks,
the oscillation with frequency $\omega_1$ is trapped between the radius, $r_{\rm c}$, where $\omega_1=(\Omega-\kappa)_{\rm c}$ 
(the subscript c denotes the value at $r_{\rm c}$) and the radius, $r_{\rm t}(> r_{\rm c})$, where the disk is truncated
by the tidal effects (see figure 1).
This oscillation has $E_1/\omega_1<0$, since $\omega_1-m_1\Omega=\omega_1-\Omega<0$ 
in the region where the oscillation dominantly exists.

\begin{figure}
\begin{center}
    \FigureFile(80mm,80mm){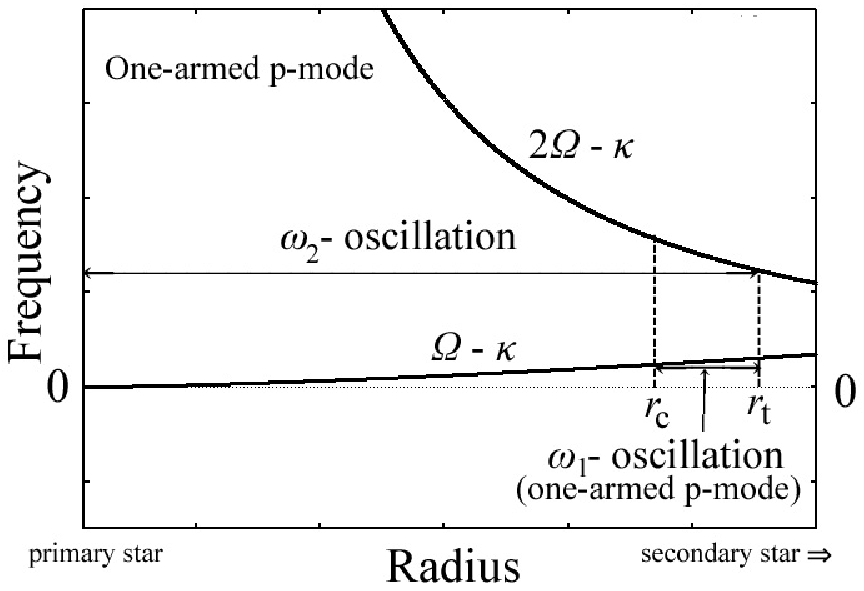}
\end{center}
\caption{Schematic diagram showing frequencies and propagation regions of the $\omega_1$- and $\omega_2$-oscillations
in the case where the $\omega_1$-oscillation is one-armed low-frequency p-mode oscillation.
The scales of coordinates are arbitrary, and are not linear.
The one-armed p-mode oscillation is trapped between $r_{\rm c}$ and $r_{\rm t}$.
The inside is the evanescent region.
The $\omega_2$-oscillation can propagate inside $r_{\rm t}$.
See figure 2 in Kato (2013), but notice that 
the signs of $\omega_2$ and $m_2$ in this paper are adopted to be opposite to those 
in paper I.
Because of this change, all the curves related to $\omega_2$-oscillations in this paper are 
line-symmetric with respect to the horizontal line of $\omega=0$  to those of figure 2 in Kato (2013).
}
\end{figure}
 
Next, let us consider a p-mode oscillation with $\omega_2$ and $m_2$.
Here, $\omega_2$ and $m_2$ are taken to be positive, and $m_2$ is found later to be 2, i.e., $m_2=2$.
The dispersion relation shows that one of the propagation region of the oscillation is specified by $\omega_2-m_2\Omega<-\kappa$.
Since $2\Omega-\kappa$ is positive and decreases outwards, the propagation region of the oscillation is bound outside.
The outer edge of the propagation region is taken to be $r_{\rm t}$  
(see figure 1) (see Kato 2013 for more discussion).
This oscillation has $E_2/\omega_2<0$, since $\omega_2-m_2\Omega<0$.

In summary, we have $(E_1/\omega_1)(E_2/\omega_2)>0$.
Furthermore, the propagation regions of the $\omega_1$- and $\omega_2$-oscillations are overlapped.
Thus, the above set of oscillations are excited 
if the set of resonant conditions, $\omega_1+\omega_2+\omega_{\rm D}=0$ and $m_1+m_2+m_{\rm D}=0$, are realized inside the disk.
These resonant conditions are summarized as
\begin{equation}
   (\Omega-\kappa)_{\rm c}+\biggr[-(1+m_{\rm D})\Omega-\kappa\biggr]_{\rm t}+m_{\rm D}\Omega_{\rm orb}=0,
\label{5.1}
\end{equation}
where $m_1=1$ and $m_2=-(1+m_{\rm D})$ are used.
For simplicity, if the difference between $\Omega$ and $\kappa$ is neglected, 
the above equation leads to
\begin{equation}
       \Omega_{\rm t}=\frac{m_{\rm D}}{2+m_{\rm D}}\Omega_{\rm orb}.
\label{5.2}
\end{equation}
For $m_{\rm D}=-3$,\footnote{
It is noted that the formal expression for resonance in this paper, i.e., $\omega_1+\omega_2+\omega_{\rm D}=0$, is
different from that of Kato (2013), i.e., $\omega_2=\omega_1-\omega_{\rm D}$.
This causes differences of values of $\omega$'s and $m$'s in this paper from those in Kato (2013), but this is formal 
and there is no difference in real results.
}
the resonant condition is satisfied by taking $r_{\rm t}$ so that $\Omega_{\rm t}$ : $\Omega_{\rm orb} =$ 3 : 1 is realized.
This can be really expected inside the disk.
The one-armed oscillation excited is the cause of superhump (Osaki 1985).

\subsection{Excitation of Low-Frequency Corrugation Mode}

Let us show that an another set of oscillations is excited in the case where $\Omega_{\rm t}$ : $\Omega_{\rm orb} =$ 3 : 1
is realized.
We consider the corrugation mode (i.e., the mode of tilt or warp with a slow frequency) as the $\omega_1$-oscillation 
(for classification of oscillation
modes in disks, see, for example, Kato 2001 and Kato et al. 2008).
The mode is characterized by $m_1=1$ and $n_1=1$, where $n_1$ is the node number in the vertical direction of the 
radial component of displacement vector, $\xi_r(\mbox{\boldmath $r$},t)$.
The vertical component of the displacement, $\xi_z(\mbox{\boldmath $r$},t)$, has no node in the vertical direction in the case of $n=1$, 
and the disk plane at a given radius oscillates up and down in the vertical direction with the same phase.

The dispersion relation shows that the propagation region of the low-frequency corrugation mode is specified by 
$\omega-\Omega<-\Omega_\bot$, 
where $\Omega_\bot(r)$ is the vertical epicyclic frequency, and is larger than $\Omega(r)$ in the tidally deformed disks, i.e., 
$\Omega_\bot>\Omega$, since
the restoring force in the vertical direction increases by the gravitational force of the secondary star.
Since $\Omega-\Omega_\bot<0$ and its absolute value increases outwards to the secondary star,
the propagation region of the oscillation is inside the radius $r_{\rm t'}$ specified by $\omega_1=(\Omega-\Omega_\bot)_{\rm t'}$,
where the subscript ${\rm t'}$ denotes the value at $r_{\rm t'}$.
The frequency of the oscillation is negative, i.e., $\omega_1<0$ (see figure 2).
This oscillation has $E_1/\omega_1<0$, since in the propagation region $\omega_1-\Omega<0$.

\begin{figure}
\begin{center}
    \FigureFile(80mm,80mm){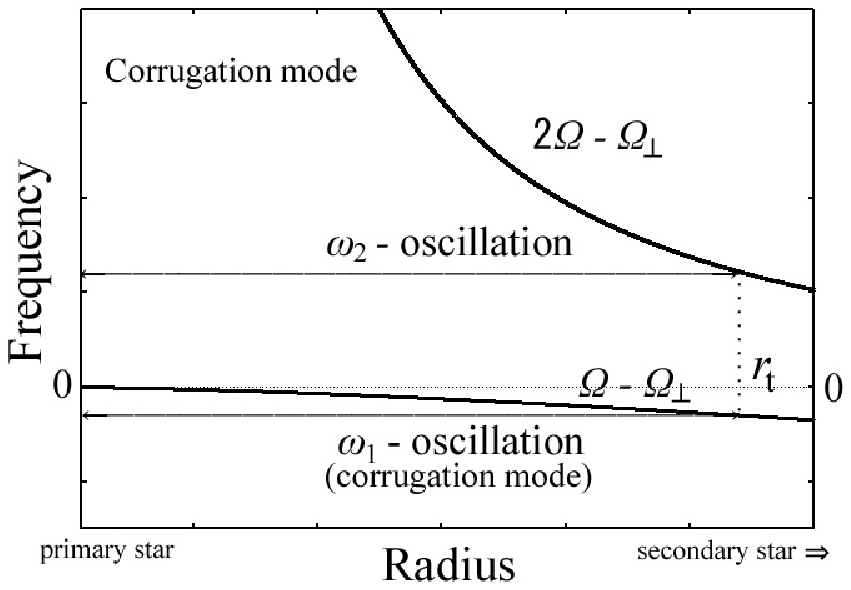}
\end{center}
\caption{Schematic diagram showing frequencies and propagation regions of the $\omega_1$- and $\omega_2$-oscillations
in the case where the $\omega_1$-oscillation is one-armed low-frequency corrugation mode.
The scales of coordinates are arbitrary, and are not linear.
The corrugation wave can propagates all region inside $r_{\rm t}$.
The $\omega_2$-oscillation can also propagate inside $r_{\rm t}$.
}
\end{figure}

As the $\omega_2$-oscillation we adopt here the vertical p-mode oscillation with $m_2=2$ and $n_2=1$.
So far, we did not mention about a resonant condition among $n_1$, $n_2$, and $n_{\rm D}$.
Unless they satisfy a certain relation, however, 
the coupling term, $\mbox{\boldmath $C$}$, vanishes and there is no resonant coupling.
In this paper a plane-symmetric tidal force is implicitly considered, which corresponds to $n_{\rm D}=0$.
Hence, for resonant coupling to occur, the $\omega_2$-oscillation must have $n_2=1$ in the vertically isothermal disks.\footnote{
In the vertically isothermal disks, the $z$-dependence of displacement vectors is expressed by use of the Hermite polynomials, and
the orthogonality of oscillation modes are classified by the index of the Hermite polynomials.
}
The dispersion relation shows that one of the propagation region of the vertical p-mode oscillation with $n_2=1$ 
is specified by $\omega_2-m_2\Omega<-\Omega_\bot$.
Now, we take $m_2=2$.
Then, since $2\Omega-\Omega_\bot$ is positive and decreases outward, the propagation region of oscillation with $\omega_2$ is 
inside the radius $r_{\rm t''}$ specified by $\omega_2=(2\Omega-\Omega_\bot)_{\rm t''} >0$,
where the subscript t'' denotes the value at $r_{\rm t''}$.
This oscillation has $E_2/\omega_2<0$, since in the propagation region $\omega_2-m_2\Omega$ is negative.
So far, the relation between $r_{\rm t'}$ and $r_{\rm t''}$ is not mentioned. 
We suppose that $r_{\rm t'}\sim r_{\rm t''}\sim r_{\rm t}$, but
careful considerations are necessary to have a reliable relation.
Here, we proceed without considering this problem in detail.

Since $(E_1/\omega_1)(E_2/\omega_2)>0$, the above set of oscillations is excited if their propagation regions are spatially
overlapped and the resonant conditions,
$\omega_1+\omega_2+\omega_{\rm D}=0$ and $m_1+m_2+m_{\rm D}=0$, are satisfied in the overlapped region.
The resonant conditions are summarized as
\begin{equation}
      (\Omega-\Omega_\bot)_{\rm t'}+(2\Omega-\Omega_\bot)_{\rm t''}+m_{\rm D}\Omega_{\rm orb}=0.
\label{5.3}
\end{equation}
The radius determined by this relation depends little on $r_{\rm t'}$ and we have
\begin{equation}
        \Omega_{\rm t''}+m_{\rm D}\Omega_{\rm orb}=0.
\label{5.4}
\end{equation}
This is satisfied for $m_{\rm D}=-3$ at the radius of $\Omega_{\rm t''}$ : $\Omega_{\rm orb} = 3$ : 1.
In summary, we have $\omega_1\sim (\Omega-\Omega_\bot)_{\rm t'}<0$ with $m_1=1$ and $\omega_2\sim(2\Omega-\Omega_\bot)_{\rm t''}>0$
with $m_2=2$. 
The signs of $\omega_1$ and $m_1$ are opposite, meaning that the oscillation is a retrograde wave.

It is interesting that oscillations which slowly progrades (one-armed p-mode oscillation) and slowly retrogrades 
(corrugation wave) are both excited when the 3 : 1 resonance is realized. 
In dwarf novae, negative superhumps are often observed in addition to (positive) superhumps.
The period of the negative superhumps is slightly shorter than the orbital period, different from those of (positive) superhumps.
Furthermore, different from (positive) superhumps which are observed only in superoutburst, negative superhumps are observed 
in superoutburst (e.g., Ohshima et al. 2012), as well as in quiescent. 
The standard interpretation of the nagative superhumps is that the disks are tilted with retrograde precession and
the gas stream from a secondary star hits the different part of the tilted disks (Wood et al. 2011).
By analyzing in detail the observational data of V1504 Cygni by the Kepler, Osaki and Kato (2013) show that the light curve of the
negative superhump in superoutburst is different from that in quiescent state, and suggest a model concerning the origin 
of the negative superhump in superoutburst.
Our results that corrugation waves, which slowly retrograde,  can  be excited if the 3 : 1 resonance is realized suggest that 
the negative superhumps in superoutburst may be related to excitation of the corrugation waves.
 
\section{Discussion}
 
We have shown that in deformed disks where the deformation is maintained by some external force or it 
has a large amplitude compared with those of disk oscillations, a set of oscillations 
satisfying 
\begin{equation}
        \biggr(\frac{E_1}{\omega_1}\biggr)\biggr(\frac{E_2}{\omega_2}\biggr)>0
\label{D.1}
\end{equation}
are excited on the disks, 
if they are eigen-mode oscillations in disks and satisfy resonant conditions, i.e., equations (\ref{2.3'}).
This problem was already examined by paper I, but there were some points to be improved in their formulation and in interpretation of 
the results.
This is the reason why we revisited to this problem.
The original purpose of considering this excitation mechanism was to suggest a possible cause of high-frequency quasi-periodic
oscillations (HF QPOs) observed in low-mass X-ray binaries (LMXBs).
On the other hand, the objects that definitely have both deformed disks and oscillatory phenomena are dwarf novae.
Hence, as an example of applications of the resonant excitation process, we have demonstrated in section 6 
how the wave-wave resonant processes can describe the tidal instability and superhump in dwarf novae.

The reason why the instability condition has such a simple form as equation (\ref{D.1})  
comes from the commutative relation of the coupling terms, i.e., equations (\ref{3.3}) and (\ref{3.3'}).
These relations are quite general. 
Even in hydromagnetic systems these relations exist, which will be shown in a subsequent paper.


It should be noticed that for $\omega_1$- and $\omega_2$-oscillations to grow, the disk deformation must be externally maintained
as in the case of tidal deformation, or has a sufficiently large amplitude compared with the oscillations.  
The HF QPOs in black-hole LMXBs appear only in the very high state (the steep power-law state). 
If the wave-wave resonant process considered here is the cause of HF QPOs, it suggest that the disks in the very high state are deformed 
from steady axisymmetric state.
Really, there are  numerical simulations which show that in the very high state a thick torus is formed 
in the innermost part of the disk and a spiral pattern appears there 
(Machida \& Matsumoto 2008).

\bigskip
The auther thanks Y. Osaki for valuable discussions on negative superhump and A.T. Okazaki for discussions 
at the stage where paper I was written.


\bigskip
\leftskip=20pt
\parindent=-20pt
\par
{\bf References}
\par
Ferreria, B.T., \& Ogilvie, G.I. 2008, MNRAS, 386, 2297 \par
Kato, S. 2001, PASJ, 53, 1\par
Kato, S. 2004, PASJ, 56, 905 \par 
Kato, S. 2008, PASJ, 60, 111 \par
Kato, S. 2011, PASJ, 63, 617 \par
Kato, S. 2012, PASJ, 64, 139\par
Kato, S. 2013, PASJ, 65, 56\par
Kato, S., Fukue, J., \& Mineshige, S. 2008, Black-Hole Accretion Disks --- Towards a New paradigm --- 
  (Kyoto: Kyoto University Press), chap. 11 \par
Kato, S., Okazaki, A.T.,\& Oktariani, F. 2011, 63, 363 (paper I) \par
Lubow, S. H. 1991, ApJ, 381, 259 \par
Lubow, S. H. 1992, ApJ, 398, 525 \par
Machida, M. \& Matsumoto, R. 2008, PASJ, 60, 613 \par
Ohshima, T., Kato, T., Pavlenko, E.P., et al. 2012, PASJ, 64, L3  \par
Oktariani, F., Okazaki, A.T., \& Kato, S. 2011, PASJ, 63, ??? \par
Osaki, Y. 1985, A\&A, 144, 369 \par 
Osaki, Y. 1996, PASP, 108, 390 \par
Osaki, Y., \& Kato, T. 2013, PASJ, 65, 50 \par 
Whitehurst, R. 1988, MNRAS, 232, 35 \par
Wood, M.A., Still, M.D., Howell, S.B., Cannizzo, J.K., \& Smale, A.P. 2011, ApJ, 741,105\par

\end{document}